\documentclass[prd,showpacs,showkeys,floatfix,nofootinbib,
               preprint,12pt,tightenlines,fleqn]{revtex4}

\usepackage{amsmath,amssymb,graphicx,dcolumn}

\begin{document}
\noindent Phys. Rev. D 77, 117901 (2008)
\hfill arXiv:0806.4351 [hep-ph] 
\newline\vspace*{0.5cm}
\title[Addendum: Ultrahigh-energy cosmic-ray bounds...]
      {Addendum: Ultrahigh-energy cosmic-ray bounds on nonbirefringent
       modified-Maxwell theory\vspace*{.125\baselineskip}}
\author{F.R.~Klinkhamer}
\email{frans.klinkhamer@physik.uni-karlsruhe.de}
\affiliation{\mbox{Institute for Theoretical Physics, University of Karlsruhe (TH),}\\
76128 Karlsruhe, Germany}
\author{M.~Risse}\email{risse@physik.uni-wuppertal.de}
\affiliation{\mbox{University of Wuppertal, Physics Department,}\\
             Gau\ss stra\ss e 20, 42097 Wuppertal, Germany\\}

\begin{abstract}
\vspace*{2.5mm}\noindent
Nonbirefringent modified-Maxwell theory,
coupled to standard Dirac particles, involves nine dimensionless
parameters, which can be bounded
by the inferred absence of vacuum Cherenkov radiation for
ultrahigh-energy cosmic rays (UHECRs). With selected UHECR events,
two-sided bounds on the eight nonisotropic parameters are obtained
at the $10^{-18}$ level, together with an improved one-sided bound on
the single isotropic parameter at the $10^{-19}$ level.
\end{abstract}

\maketitle

In Ref.~\cite{KlinkhamerRisse2008},
ultrahigh-energy-cosmic-ray (UHECR) bounds have been given for the nine
Lorentz-violating ``deformation parameters'' of nonbirefringent
modified-Maxwell theory coupled to standard Dirac particles,
where the parameters were restricted to a particular domain.
In this addendum, we obtain corresponding results for
two sets of nonisotropic parameters \emph{outside} this domain
(the two sets are, respectively, parity-odd and parity-even).
These new bounds are essentially \emph{two-sided},
whereas an improved bound
on the single isotropic parameter remains one-sided.
For convenience, the final bounds will be presented
in terms of the widely-used standard-model-extension (SME)
parameters \cite{ColladayKostelecky1998,KosteleckyMewes2002}.

\begin{table*}[b]
\vspace*{5mm}
\begin{center}
\caption{Selected UHECR events with energies above $57\;\text{EeV}$ as
recorded by the Pierre Auger Observatory over the period 1 January 2004 to 31
August 2007 \cite{Abraham-etal2007}, together with an additional
$320\;\text{EeV}$ event from the Fly's Eye detector \cite{Bird-etal1995} and
a $210\;\text{EeV}$ event from the AGASA array \cite{Hayashida-etal1994}.
Shown are the arrival time (year and Julian day), the primary energy $E$, and
the arrival direction with right ascension $\text{RA}$ and declination
$\text{DEC}$. The estimated errors for the Auger events
\cite{Abraham-etal2007} are a $25\,\%$ relative error on the energy and a
$1^\circ$ error in the pointing direction (the errors for the Fly's Eye and
AGASA events are of the same order, possibly somewhat larger
\cite{Bird-etal1995,Hayashida-etal1994}).
\vspace*{5mm}} \label{tab-highE-Auger-events}
\renewcommand{\tabcolsep}{.5pc}    
\renewcommand{\arraystretch}{.9}   
\begin{tabular}{lrrrr|lrrrr}
\hline\hline
  Year & Day &  E\! [EeV]  & RA\! [deg] &      DEC\! [deg] &
\;Year & Day &  E\! [EeV]  & RA\! [deg] &      DEC\! [deg]\\
\hline
$ 1991 $&$ 288 $&$ 320 $&$  85.2 $&$   48.0  $& 
$\; 2006 $&$  81 $&$  79  $&$ 201.1 $&$  -55.3 $\\
$   1993 $&$ 337 $&$ 210 $&$  18.9 $&$   21.1  $& 
$\; 2006 $&$ 185 $&$  83  $&$ 350.0 $&$    9.6 $\\
$   2004 $&$ 125 $&$  70  $&$ 267.1 $&$  -11.4 $&
$\; 2006 $&$ 296 $&$  69  $&$  52.8 $&$   -4.5 $\\
$   2004 $&$ 142 $&$  84  $&$ 199.7 $&$  -34.9 $&
$\; 2006 $&$ 299 $&$  69  $&$ 200.9 $&$  -45.3 $\\
$   2004 $&$ 282 $&$  66  $&$ 208.0 $&$  -60.3 $&
$\; 2007 $&$  13 $&$ 148  $&$ 192.7 $&$  -21.0 $\\
$   2004 $&$ 339 $&$  83  $&$ 268.5 $&$  -61.0 $&
$\; 2007 $&$  51 $&$  58  $&$ 331.7 $&$    2.9 $\\
$   2004 $&$ 343 $&$  63  $&$ 224.5 $&$  -44.2 $&
$\; 2007 $&$  69 $&$  70  $&$ 200.2 $&$  -43.4 $\\
$   2005 $&$  54 $&$  84  $&$  17.4 $&$  -37.9 $&
$\; 2007 $&$  84 $&$  64  $&$ 143.2 $&$  -18.3 $\\
$   2005 $&$  63 $&$  71  $&$ 331.2 $&$   -1.2 $&
$\; 2007 $&$ 145 $&$  78  $&$  47.7 $&$  -12.8 $\\
$   2005 $&$  81 $&$  58  $&$ 199.1 $&$  -48.6 $&
$\; 2007 $&$ 186 $&$  64  $&$ 219.3 $&$  -53.8 $\\
$   2005 $&$ 295 $&$  57  $&$ 332.9 $&$  -38.2 $&
$\; 2007 $&$ 193 $&$  90  $&$ 325.5 $&$  -33.5 $\\
$   2005 $&$ 306 $&$  59  $&$ 315.3 $&$   -0.3 $&
$\; 2007 $&$ 221 $&$  71  $&$ 212.7 $&$   -3.3 $\\
$   2005 $&$ 306 $&$  84  $&$ 114.6 $&$  -43.1 $&
$\; 2007 $&$ 234 $&$  80  $&$ 185.4 $&$  -27.9 $\\
$   2006 $&$  35 $&$  85  $&$  53.6 $&$   -7.8 $&
$\; 2007 $&$ 235 $&$  69  $&$ 105.9 $&$  -22.9 $\\
$   2006 $&$  55 $&$  59  $&$ 267.7 $&$  -60.7 $&
&&&&\\
\hline\hline
\end{tabular}
\end{center}
\end{table*}

The `Note Added in Proof' of Ref.~\cite{KlinkhamerRisse2008} used
29 UHECR events \cite{Abraham-etal2007,Bird-etal1995,Hayashida-etal1994}
which, for completeness, are listed in Table~\ref{tab-highE-Auger-events}.
From the energies and flight directions
of these 29 UHECR events, the following two--$\sigma$ bound
was obtained on the quadratic sum of the nine nonbirefringent
Lorentz-violating parameters $\alpha^l$ \cite{KlinkhamerRisse2008}:
\begin{subequations}\label{eq:vecalphabound-addendum}
\begin{eqnarray}
\vec{\alpha} &\in& D_\text{causal}^\text{(open)}\;:\;\;\; |\vec{\alpha}|^2
\equiv \sum_{l=0}^{8}\, \big(\alpha^l\big)^2 < A^2\,,
\label{eq:vecalphabound-general-addendum}\\[1mm]
A &=& 3 \times 10^{-18} \,\left(
\frac{M_\text{prim}}{56\:\text{GeV}/c^2}\right)^{2},
\label{eq:vecalphabound-avalue-addendum}
\end{eqnarray}
\end{subequations}
showing explicitly the dependence on the mass of the primary charged
particle (taken equal for all events).
There are indications \cite{Abraham-etal2007} that these UHECRs originate
predominantly from protons but, in order to be on the safe side,
we will later take the mass $M_\text{prim}$ to be equal to that of iron,
$M_\text{prim}=56\:\text{GeV}/c^2$.
Bound \eqref{eq:vecalphabound-addendum}
as well as all other bounds in this addendum are based on the
Cherenkov threshold condition (10) in Sec. II B of Ref.~\cite{KlinkhamerRisse2008}
and the reader is referred to this section, in particular, for further details.

\par
The domain used in \eqref{eq:vecalphabound-general-addendum} is defined by
\begin{equation}\label{eq:Dcausalopen-addendum}
D_\text{causal}^\text{(open)} \equiv \{ \vec{\alpha} \in \mathbb{R}^9 :\;
\forall_{\widehat{x}\in \mathbb{R}^3} \; (\alpha^0 + \alpha^j\,\widehat{x}^j+
\widetilde{\alpha}^{jk}\,\widehat{x}^j\,\widehat{x}^k)>0 \},
\end{equation}
where $\widehat{x}\equiv \vec{x}/|\vec{x}|$ denotes a unit
vector in Euclidean three-space and the traceless symmetric $3\times 3$
matrix $\widetilde{\alpha}^{jk}$ is defined in terms of the parameters
$\alpha^l$ for $l=4, \ldots , 8$ (see below).
The parameter domain \eqref{eq:Dcausalopen-addendum} allows for vacuum Cherenkov
radiation in
\emph{all} directions and, with boundaries added, is believed to constitute a
significant part of the physical domain of the theory, where, e.g., unitarity
and microcausality hold; cf. Appendix C of Ref.~\cite{KaufholdKlinkhamer2007}. It
may, nevertheless, be of interest to get bounds \emph{outside}
this domain, because modified-Maxwell theory could be only part of
the full Lorentz-noninvariant theory.

The crucial observation is that domain \eqref{eq:Dcausalopen-addendum}
shrinks to zero size in the hyperplane $\alpha^0=\widetilde{\alpha}^{jk}=0$,
so that bound \eqref{eq:vecalphabound-general-addendum} becomes ineffective there.
Still, the data from
Table~\ref{tab-highE-Auger-events} can be used to get the following two--$\sigma$
bound on the three parity-odd nonisotropic parameters in this hyperplane
\begin{eqnarray}\label{eq:two-sided-bounds123-addendum}
\hspace*{-1cm}&&
\alpha^0=\alpha^4=\alpha^5=\alpha^6=\alpha^7=\alpha^8=0 \;:\;\;\;
\sum_{j=1}^{3}\,(\alpha^j)^2 <  \Big(4 \times 10^{-18}\Big)^2 \,
\left(\frac{M_\text{prim}}{56\:\text{GeV}/c^2}\right)^{4} \,.
\end{eqnarray}
Similarly, there is a two--$\sigma$ bound on the five
parity-even nonisotropic parameters in an orthogonal hyperplane
\begin{eqnarray}\label{eq:two-sided-bounds45678-addendum}
\hspace*{-1cm}&&
\alpha^0=\alpha^1=\alpha^2=\alpha^3=0 \;:\;\;\;
\sum_{l=4}^{8}\,(\alpha^l)^2
<  \Big(4 \times 10^{-18}\Big)^2 \,
\left(\frac{M_\text{prim}}{56\:\text{GeV}/c^2}\right)^{4} \,.
\end{eqnarray}
It is, in principle,
possible to get other bounds for the eight nonisotropic parameters, but,
for the moment, bounds \eqref{eq:two-sided-bounds123-addendum}
and \eqref{eq:two-sided-bounds45678-addendum} suffice.

If only a \emph{single} parameter $\alpha^l$ for $l\in \{1,\ldots,8\}$ is
considered (all seven other  nonisotropic parameters and the isotropic
parameter $\alpha^0$ being zero), bounds \eqref{eq:two-sided-bounds123-addendum}
and \eqref{eq:two-sided-bounds45678-addendum} give a two-sided
bound on that single isolated parameter. Setting $M_\text{prim}=56\:\text{GeV}/c^2$
and showing explicitly the approximate one--$\sigma$ error, these bounds are
\begin{equation}\label{eq:two-sided-bounds12345678-addendum}
l\in \{1,\ldots,8\}\;:\;\;\; |\alpha^l| <  (2 \pm 1) \times 10^{-18}\,,
\end{equation}
for $\alpha^0=\alpha^m=0$ with $m\in \{1,\ldots,8\}$ and $m\ne l$.
Incidentally, the possibility of getting certain two-sided Cherenkov bounds
from an isotropic set of UHECR events has already been noted in Appendix C of
Ref.~\cite{KaufholdKlinkhamer2007}.

For completeness, we also give the following one-sided bound on the single
$\alpha^0$ parameter
\begin{equation}\label{eq:one-sided-bound0-addendum}
\alpha^0 <  (1.4 \pm 0.7) \times 10^{-19}\,,
\end{equation}
for $\alpha^m=0$ with $m\in \{1,\ldots,8\}$.
Bound \eqref{eq:one-sided-bound0-addendum} has been derived by setting
$M_\text{prim}=56\:\text{GeV}/c^2$ and using the $148\:\text{EeV}$ Auger event
from Table~\ref{tab-highE-Auger-events}, which has a reliable energy
calibration \cite{Abraham-etal2007}.
For the Fly's Eye event with an estimated energy of
$320\:\text{EeV}$ \cite{Bird-etal1995}, bound \eqref{eq:one-sided-bound0-addendum}
would be reduced by a factor of approximately $5$ according to
Eq.~(10) in Ref.~\cite{KlinkhamerRisse2008}.

In order to facilitate the comparison with existing laboratory bounds
and future ones, we provide a dictionary between our $\alpha^l$ (or
$\widetilde{\alpha}^{\mu\nu}$) parameters and the nonbirefringent
SME parameters defined by Eq.~(11) in Ref.~\cite{KosteleckyMewes2002}:
\begin{equation}\label{eq:alpha-parameters-dictionary-addendum}
\vec{\alpha} \equiv \left(
  \begin{array}{c}
    \alpha^0 \\
    \alpha^1 \\
    \alpha^2 \\
    \alpha^3 \\
    \alpha^4 \\
    \alpha^5 \\
    \alpha^6 \\
    \alpha^7 \\
    \alpha^8 \\
  \end{array}
\right)
 \equiv
\left(
  \begin{array}{c}
    \widetilde{\alpha}^{00} \\
    \widetilde{\alpha}^{01} \\
    \widetilde{\alpha}^{02} \\
    \widetilde{\alpha}^{03} \\
    \widetilde{\alpha}^{11} \\
    \widetilde{\alpha}^{12} \\
    \widetilde{\alpha}^{13} \\
    \widetilde{\alpha}^{22} \\
    \widetilde{\alpha}^{23} \\
  \end{array}
\right) = \left(
  \begin{array}{c}
    2\,\widetilde{\kappa}_\text{tr}\\
    -2\,(\widetilde{\kappa}_{\text{o}+})^{(23)}\\
    -2\,(\widetilde{\kappa}_{\text{o}+})^{(31)}\\
    -2\,(\widetilde{\kappa}_{\text{o}+})^{(12)}\\
    -(\widetilde{\kappa}_{\text{e}-})^{(11)}\\
    -(\widetilde{\kappa}_{\text{e}-})^{(12)}\\
    -(\widetilde{\kappa}_{\text{e}-})^{(13)}\\
    -(\widetilde{\kappa}_{\text{e}-})^{(22)}\\
    -(\widetilde{\kappa}_{\text{e}-})^{(23)}\\
  \end{array}
\right).
\end{equation}
The Cartesian coordinates employed
(cf. Sec. III A of Ref.~\cite{KosteleckyMewes2002}) are such that
the flight-direction vector $\widehat{q}$
of an UHECR primary at the top of the Earth atmosphere is given by
\begin{equation}\label{eq:qhat-addendum}
\left( \begin{array}{c}
  \widehat{q}_1 \\
  \widehat{q}_2 \\
  \widehat{q}_3
\end{array} \right)
= -
\left(\begin{array}{l}
  \sin(\pi/2-\delta)\cos\alpha \\
  \sin(\pi/2-\delta)\sin\alpha \\
  \cos(\pi/2-\delta)
\end{array}\right)\,,
\end{equation}
in terms of the celestial coordinates
$\text{RA}\equiv \alpha$ and $\text{DEC}\equiv \delta$ from
Table~\ref{tab-highE-Auger-events}.

\par
Using the dictionary \eqref{eq:alpha-parameters-dictionary-addendum},
bounds \eqref{eq:two-sided-bounds12345678-addendum} and
\eqref{eq:one-sided-bound0-addendum} give the following
two--$\sigma$ ($95\%\:\text{CL}$) bounds on the nine isolated
SME parameters of nonbirefringent modified-Maxwell theory:
\begin{subequations}\label{eq:SMEbounds-nine} 
\begin{eqnarray}
\hspace*{-5mm}
\big| (\widetilde{\kappa}_{\text{o}+})^{(ij)}
\big|_{(ij)=(23),(31),(12)} &<& 2 \times 10^{-18}\,,
\label{eq:SMEbounds-nonisotropic-odd}\\[2mm]
\hspace*{-5mm}
\big| (\widetilde{\kappa}_{\text{e}-})^{(kl)}
\big|_{(kl)=(11),(12),(13),(22),(23)} &<& 4 \times 10^{-18}\,,
\label{eq:SMEbounds-nonisotropic-even}\\[2mm]
\hspace*{-5mm}
\widetilde{\kappa}_\text{tr} &<& 1.4 \times 10^{-19}\,,
\label{eq:SMEbounds-isotropic}
\end{eqnarray}\end{subequations}
for the Sun-centered Cartesian coordinate system employed
in \eqref{eq:qhat-addendum}.
The Cherenkov bounds \eqref{eq:SMEbounds-nonisotropic-odd},
\eqref{eq:SMEbounds-nonisotropic-even}, and \eqref{eq:SMEbounds-isotropic}
are significantly stronger than the current laboratory bounds
at the $10^{-12}$, $10^{-16}$, and $10^{-7}$ levels, respectively;
see the third paragraph of Sec. V in Ref.~\cite{KlinkhamerRisse2008}
for further discussion and references.
It is to be emphasized that these Cherenkov bounds
only depend on the measured energies and flight directions
of the charged cosmic-ray primaries at the top of the Earth atmosphere.

\section*{\hspace*{-5.5mm}ACKNOWLEDGMENTS}
\noindent
FRK acknowledges the hospitality of
The Henryk Niewodnicza\'{n}ski Institute of Nuclear Physics
in Cracow, Poland, where part of this work was done,
and the help of M. Schreck with
the signs in \eqref{eq:alpha-parameters-dictionary-addendum}.
Both authors thank V.A. Kosteleck\'{y} for useful suggestions
regarding the presentation of the results of this addendum.

\end{document}